# Compact, Robust Source of Cold Atoms for Efficient Loading of a Magnetic Guide

**R.S. Conroy, Y. Xiao, M. Vengalattore, W. Rooijakkers, M. Prentiss**

Center for Ultracold Atoms,

Jefferson Laboratories,

Department of Physics, Harvard University,

Cambridge.  MA02138  U.S.A.

Contact Details:

Tel. +1-617-495-4483   Fax. +1-617-495-0416   E-mail: RConroy@fas.harvard.edu

28 August, 2003

**Abstract:**

We report a compact ($<20cm^3$), robust source for producing a bright flux of cold atoms, which can be loaded efficiently into a magnetic guide. A continuous flux of up to 8 x $10^9$ $^{87}$Rb atoms/s have been produced from this 2D+ vapor cell MOT. The flux had a divergence of 12.5 mrad and velocity could be controlled in the range 2-15 m/s. This flux was coupled continuously into a magnetic guide with high efficiency.

# Compact, Robust Source of Cold Atoms for Efficient Loading of a Magnetic Guide

**R.S. Conroy, Y. Xiao, M. Vengalattore, W. Rooijakkers, M. Prentiss**

Jefferson Laboratories,

Department of Physics, Harvard University,

Cambridge.  MA02138  U.S.A.

## Introduction

In order to probe the macroscopic quantum properties of dilute atomic gases, ultra-high vacuum conditions are required to minimize deleterious environmental interactions. This contrasts with the high background vapor pressure required to collect large numbers of cold atoms. Therefore it is desirable to have a region with high background pressure dedicated to producing a continuous flux of atoms, which can then be coupled into an ultra-high vacuum region.

This approach has been very successful for the production of cold atomic beams, in particular using Zeeman slowers [1] and vapor cells [2]. Zeeman slowers produce a large flux, but with a high mean velocity and potential issues with size, complexity, stray magnetic fields and the use of light along the atomic beam axis. To produce a cold and collimated flux, secondary cooling of the output of a Zeeman slower is required, adding further complexity. Vapor cells trade off flux for a lower mean velocity, but with similar flux density and a greater flexibility in design. It is worth noting that the density of the flux from all these sources is typically within an order of magnitude and therefore the choice of source is primarily determined by the requirements of the downstream experiment.

The goal of this work was to produce a compact, robust source for loading magnetic guides efficiently. This requires a bright source of cold atoms with a low velocity and narrow spread, qualities best matched to a vapor cell design. The three most common vapor cell designs are the two dimensional magneto-optical trap (2D MOT) [3], the 2D+ MOT [4] and the low-velocity intense source (LVIS) [5]. A 2D MOT, where a two dimensional quadrupole magnetic field is used in conjunction with optical cooling along the two orthogonal axes to the flux, is attractive when light along this axis is a disadvantage, but it does not significantly reduce or narrow the velocity distribution along the flux axis. Both the 2D+ MOT and LVIS designs use cooling and radiation force imbalance along the flux axis to produce a slow and narrow velocity distribution significantly to the several meters per second range, using a two-dimensional and three-dimensional quadrupole magnetic field respectively. To maximize the capture in velocity-volume space and the coupling efficiency of the flux from the vapor cell into the 2D magnetic field of a magnetic guide, a 2D+ MOT design was chosen.

There is significant interest in magnetic guides for a number of applications from nonlinear optics [6,7] to interferometry [8,9] and the generation of degenerate states [10,11]. The design of the magnetic guide used here has been previously reported and shown to be an efficient and flexible design [12]. Previously our guides have been loaded in-situ from a hot background and consequently had a limited trap lifetime and could not be used to continuously load a magnetostatic trap. Elsewhere a number of other approaches have been used to load magnetic guides [13], though issues with heating, poor efficiency and lack of optical access have limited their widespread implementation. More recently there has been a concerted effort to efficiently inject a continuous flux into a magnetic guide using low velocity, bright beams to address these applications [14,15].

In this article we report the construction and optimization of a simple, compact, robust 2D+ vapor cell MOT with efficient coupling of the flux into a magnetic guide. This scheme provides a model for loading atoms in a dark state into a magnetic guide with the future goal of investigating transport and evaporative cooling of the atoms within the guide. The primary goal of this work therefore is to inject as high a starting density into the guide as possible with a low, controllable velocity and narrow spread.

## Setup

Figure 1 illustrates the experimental setup. A 25x25x67mm standard quartz cell was divided in two equal sections by a 25x25x3mm gold mirror. In the center of the mirror a 1.25mm diameter hole was drilled to connect the vapor cell in the front half of the cell with the magnetic guide in the rear half, each half with a volume of less than 20cm$^3$.

Two rubidium getter ovens provided a simple, compact, controllable method for varying the background rubidium density between $10^7$-$10^{10}$ atoms/cm$^3$ in the front half of the cell. The base pressure of the system was 3 x $10^{-9}$ mbar created by a 100 liter/s turbo pump, and rose to 8 x $10^{-8}$ mbar on the UHV side of the cell when the getters were operating at maximum current. To form the 2D+ MOT, a two dimensional quadrupole magnetic field was created by four coils wrapped around the cell. These coils also provided direct heating of the vapor cell, preventing rubidium from condensing on the cell walls and enabling higher vapor pressures. The two transverse, circularly-polarised beams were retro-reflected using right-angle prisms fixed to the cell using optical cement, making the design more simple and robust. The third beam along the axis of the atomic flux, with a shadow mask for the region of the hole was reflected by the gold mirror to provide optical molasses in the longitudinal direction. A separate 1mm-diameter "pushing beam" co-propagating with the flux was overlapped with hole to provide a way of controlling the flux through the hole. A counter-propagating "retarding beam" of the same size was also used for flux control. The beam size was chosen to be smaller than the hole size to avoid interaction with the hole and to extract only the coldest atoms from the 2D+ MOT. Using separate pushing and retarding beams enhances the control of the flux produced over conventional LVIS and 2D+ MOT designs through the choice of the size, intensity and detuning of these beams. Both beams were linearly polarized to aid manipulation and measurement, however to reach much lower velocities or if a magnetic field or field gradient is used along the flux axis then circularly-polarised light can be advantageously used.

The magnetic guide was composed of four m$\mu$-metal foils (25x7x0.5mm) with 3mm spacing between them and thirty turns of kapton wire around them. A 0.5mm thick gold-coated silicon wafer was cut to size and placed on top to complete the magneto-optical design [12]. The surface of the wafer was 3mm below the center of the hole leading from the vapor cell, illustrated in the inset of Figure 1. The flux was measured experimentally via on-resonance absorption 5mm and 15mm after the hole on the magnetic guide side of the experiment. The visible length of the guide was limited to 20mm by mounting of the components.

## Vapor Cell Optimization

The flux out of the vapor cell can never be greater than the loading rate into the MOT, though it can of course be much less. The loading rate into the MOT can be increased by increasing the capture velocity and capture volume of the MOT, and/or by increasing the background vapor in the cell. Not all atoms successfully loaded into the MOT can be transported out of the cell, so optimization of the MOT loading rate does not necessarily optimize the flux. For example, though the loading rate increases linearly with background pressure, the flux out does not because collisions due to the background gas prevent atoms loaded into the MOT from being transferred out of the cell. In addition, for a given density in the MOT, the flux out increases with increasing velocity, but at high velocities the atoms are transported out of the 2D+ MOT region faster than they are captured, so the MOT density and beam density decreases with increasing flux velocity at high velocities. Finally, the loading into the magnetic guide may not be optimized at the parameters that provide the largest flux out of the vapor cell. For our application it is important to optimize the parameters for maximum flux density.

The first parameter which can be easily considered is the background vapor pressure. For a flux velocity of 10m/s, background collisions would start to limit the flux density above a background density of $1 \times 10^{10}$ atoms/cm$^3$. The optimum flux density for this cell was obtained at background pressures of $1 \times 10^{10}$ atoms/cm$^3$ in agreement with the measured flux velocity of 11m/s.

Experimentally, there is a trade-off between capture volume and capture velocity for a fixed laser power. Simple MOT modeling indicates that the loading efficiency can be maximized when the intensity of each beam is approximately ten times the saturation intensity. For the 300mW available from the tapered amplifier used in these experiments, this equates to beam diameters of 35-23mm, matching the size of the cell windows. Figure 2 illustrates how the flux density varies with beam size and intensity with the other experimental parameters optimized. The transverse light intensity does not significantly affect the flux velocity, but the density peaks around a transverse beam intensity of 11mW/cm$^2$. The longitudinal light intensity, including the pushing beam, clearly controls the flux velocity, with a peak in the density similar to the transverse beams. From Figure 2b) the flux scales as expected as the two-thirds power of the diameter of the beams. This result indicated we have efficient capture over the whole cell size to within two millimeters of the cell wall, that we are not density limited and that collisions with the background are not a significant source of loss for the cold atomic flux.

Optimization of the magnetic field determines the ideal laser detuning and hence the maximum capture velocity from the hot background vapor. If the magnetic field gradient is too small, then atoms will not be sufficiently damped to produce an intense atomic beam, whereas too high a gradient limits the distance out to which capture will take place. To maintain good damping of atoms in the magneto-optic trap, the gradient of the magnetic field needs to be at least 10G/cm. The optimum laser detuning is typically around two linewidths, limiting the capture radius to 8mm for these parameters. However the maximum capture velocity can be increased by using power broadening to red-shift the optimum laser detuning by several linewidths. By setting the transverse beam intensities to 11mW/cm, the flux density was maximized with a detuning of 30MHz and a magnetic field gradient of 21.5G/cm, giving efficient capture throughout the cell.

Although not implemented here, the laser frequency can also be scanned rapidly to broaden the velocity space which can be captured.

The hole size required for coupling the atoms out of the vapor cell can be found by considering the diameter of the cold atoms confined in the radial magnetic field. It is desirable that the hole size should match the diameter of the MOT for efficient coupling and to minimize hot background collisions beyond the hole.

This was confirmed experimentally by offsetting the trap center from the axis of the hole to form a stable MOT. The resulting cloud was found to have a diameter of 1.4mm and a length of 21mm at a magnetic field gradient of 20G/cm, which was reduced to a diameter of 1.2mm and a length of 19mm at 27G/cm. The flux-optimized values, used subsequently, were measured to be 23G/cm and a laser detuning of 28MHz [$\Gamma$=-5.5].

The choice of hole size is also affected by the transverse temperature of the cold atoms because of divergence. However as the longitudinal velocity is approximately two orders of magnitude greater than the transverse velocity and the solid angle filtering due to the hole size ($\phi$=1.25mm x 3mm), the angular spread was expected to be small. The divergence of the flux after the hole was measured to be 12.5mrad, equating to a transverse temperature of 80μK. Increasing the hole size increases the divergence of the beam by permitting hotter atoms to escape. There is no significant gain in cold flux when the diameter of the hole significantly exceeds that of a stable MOT.

The scalability of this vapor cell design has also been considered. Increasing the length of the cell increases the path length for an atom to leave the vapor cell, limiting the background density that can be used before collisions limit the flux density. There will also be an increase in the time an atom spends on axis with the hole where it will experience acceleration, giving a higher average longitudinal velocity, while giving a slightly smaller transverse velocity. The longer cell length will also increase the transverse cooling time, permitting a purely 2D MOT to work. Increasing the transverse size of the cell will increase the potential capture volume and decrease the effect of the walls, but also decrease the magnetic field gradient leading to a larger and hotter flux of cold atoms. Figures 2, 3 and our Monte Carlo modeling illustrate these points well and indicates that ideally we our cell size should be approximately 50% larger than the one used, given our laser power and magnetic guide, yielding up to five times more flux.

## Flux Control

Figure 3 shows how the flux density varies with the flux velocity. The flux velocity was controlled by changing the intensity ratio between the pushing and retarding beams from 1:0 to 1:1, corresponding to a reduced acceleration of the atoms in the region of the hole and a decrease in the velocity. The measured flux density achieved for different background rubidium densities is plotted, and it is worth noting the cold atomic flux density is approximately that of the background hot atoms and there is a linear increase in flux density with background pressure, indicating that background collisions are not limiting the flux. When optimized, the maximum turn-key flux density from the vapor cell was measured to be $5.8 \times 10^8$ atoms/cm$^3$ with a velocity of $11 \pm 1.5$m/s, and a maximum flux of $8 \times 10^9$ atoms/s at a velocity of 15m/s. This is approximately 50% of the optimal flux of $1.2 \times 10^{10}$ atoms/s from a maximum capture velocity of 52m/s from a background rubidium vapor pressure of $1 \times 10^{10}$ atoms/cm$^3$ in a 20cm$^3$ capture volume with a fill time of 5ms. The loss in efficiency is attributed to the anisotropy of the diode laser's intensity profiles particularly for the transverse beams close to the hole where there is mismatch in the light intensities due to aperturing of beams at the edge of the prisms as well as a small mismatch in their positioning, causing transverse heating and deflection. There is also a loss factor associated with the hot background atoms exiting through the hole due to pumping. Ideally the flux density should increase as the pushing force and velocity decrease. However a sharp drop is observed because of the increased transit time out of the cell and the increased importance of transverse velocity and any light force imbalances. If a lower velocity is desirable, then a pair of far-detuned moving molasses system could be used instead.

By pulsing the longitudinal light on axis with the hole, it was found that a pulsed flux could be obtained, as shown in Figure 2a). The peak pulsed flux was four times that of the continuous flux with a repetition rate of up to 150Hz and up to $1 \times 10^7$ atoms/pulse. The pulsed operation arises because the small cell length and hole size offers extremely poor output coupling for a purely 2D MOT when the pushing beam is switched off. When the light is switched on, the reservoir of transversely cooled atoms can then be pushed out through the hole by the pushing beam. A pulsed flux is an attractive alternative for loading a magnetic guide because it can reduce the amount of stray light causing decoherence and permit more efficient optical pumping into a trapped magnetic state. These efficiencies potentially recover the loss in overall flux and a continuous flux in the guide can be recovered over time if the longitudinal velocity spread is considered.

The flux velocity is an important experimental parameter in a magnetic guide because too low a velocity will limit the flux available from the vapor cell source and increase the potential for heating in a continuous beam with an axial temperature gradient; too high a velocity will require an excessively long magnetic guide for downstream experiments. Control of the flux velocity was investigated using two methods: by varying the intensity of the counter-propagating "retarding beam" and by frequency detuning it from the 28MHz red-detuned "pushing beam". Figure 4 illustrates that below the saturation intensity, any counter-propagating light has little effect on the velocity. However as the intensity approaches that of the beam propagating in the direction of the flux, the velocity of the atoms decreases to 5m/s.

The effect of detuning the counter-propagating light was also investigated. A beam matched to the intensity of the propagating beam was detuned in the range –6 to 4 MHz using two 40MHz acousto-optic modulators. Accurate velocity control from 2.5 to 7m/s could be achieved in this way. In principle must lower velocities can be obtained using this moving molasses approach [10], however with the relatively high background pressure used here, there was a 55% decrease in flux as the velocity was decreased to 2.5m/s.

## Loading a Magnetic Guide

The choice of hole size determined the spacing of the foils used in the magnetic guide (3mm). Modeling of the magnetic fields indicates that the hole size needs to be smaller than half the spacing of the foils for efficient transfer between the geometries.

The inset in Figure 1 illustrates how the flux from the vapor cell can be used to load a magneto-static trap. The first goal of the loading process is to separate the flux of cold atoms from the flux of hot background atoms, which also exit the vapor cell through the hole. By choosing the current directions in the four-foils used, the magnetic field generated reinforced the 2D quadrupole field of the vapor cell MOT. The height of the magnetic field minimum was adjustable by changing the currents around the foils to produce an axis for the guided flux 0-3mm below the axis of the flux exiting the vapor cell. By choosing the lengths and curvature of the mμ-metal foils correctly, edges effects can be minimized. In addition, the absence of repumper light in the region around and through the hole reduces the forces on the atomic flux as it enters the magnetic field above the above the guide. A short (5mm long) transverse cooling and repumper section 5mm after the hole provides sufficient momentum for the cold atoms to make the transition to the magnetic field minimum of the guide but not the residual hot atoms. The cooling beam was chosen to be large enough to provide sufficient time for the cold flux to receive enough transverse momentum to relocate and be cooled in the axis of the guide, but short enough not to significantly alter the trajectory of hot atoms also exiting the hole. This beam was derived from the main cooling light of the vapor cell and therefore had the same detuning and the transfer was optimized with an intensity of 10 $I_{sat}$.

The temporal sequence normally used to efficiently load a magnetic trap from a magneto-optical trap cannot readily or efficiently be employed here because of the need to define a quantisation axis and optically pump the atoms into a low-field seeking state. Using a pulsed flux would remove these limitations, however it is not clear the resulting flux would be greater than simply employing a depumping beam immediately after the transverse cooling section. By pumping the atoms into the F=1 state, approximately 33% of the atoms will end up in the weak-field seeking $m_f$ = -1 sublevel and be guided by the magnetic guide.

The continuous loading efficiency of the flux into the magnetic guide was investigated with no counter-propagating beam to avoid issues with light scattering. The size of the pushing beam as mentioned earlier was chosen to be smaller than the hole size to push out only the coldest atoms and to avoid aperturing, diffraction and light scattering effects from the hole. The maximum magnetic field gradient that could be produced by the guide was 100G/cm with the minimum 2mm below and 0.4mm transversely offset from the flux axis through the hole. This spatial separation between the flux coming through the hole and the axis of the guide minimizes the effect of the pushing light on the atoms in the magnetic guide. Due to spatial constraints no axial magnetic holding field was employed to prevent spin-flips, however because of the short distance over which detection could be made these were not considered to be a major source of loss.

The transfer efficiency into the guide was measured using a second on-resonant absorption beam 5mm after the transverse cooling section. Along the original flux axis from the vapor cell, this beam measured a depletion of 85%. When adjusted to the height of the magnetic field minimum of the guide, 40% of the original continuous flux of atoms was detected, with a density of $2\times10^9$ atoms/cm$^3$. This measured efficiency is several times greater than other methods of loading magnetic guide [15,16] and indeed is higher than expected in part because the detection zone was not sufficiently far from the loading zone to guarantee only the magnetically trapped atoms were detected. From an approximate calculation however these background atoms contribute no more than 35% to the measured signal. Although no significant heating of the flux or change in the longitudinal velocity was observed, further quantitative measurements of the temperature could not be made because of experimental limitations.

## Conclusions

We have demonstrated a compact, turnkey source of cold atoms for efficiently loading a magnetic guide. A maximum continuous flux of $8 \times 10^9$ $^{87}$Rb atoms/s with a velocity of $15 \pm 1.5$m/s, diameter of 1.25mm and divergence of 12.5mrad has been produced from a 2D+ vapor cell MOT. This flux was loaded efficiently into a magnetic guide.

This technique can be scaled and made more efficient to yield an order of magnitude more atoms to provide an attractive route to generating a bright, robust source of cold atoms. After this proof of concept, we are developing a longer magnetic guide structure into which the atoms can be loaded and characterized more completely, in particular the most efficient route to controlling the velocity and preparing the flux in a magnetically trappable state.

## Acknowledgements

This work was funded by the NSF grants for the Center for Ultracold Atoms (PHY-0071311) and MURI (19-00-0007).

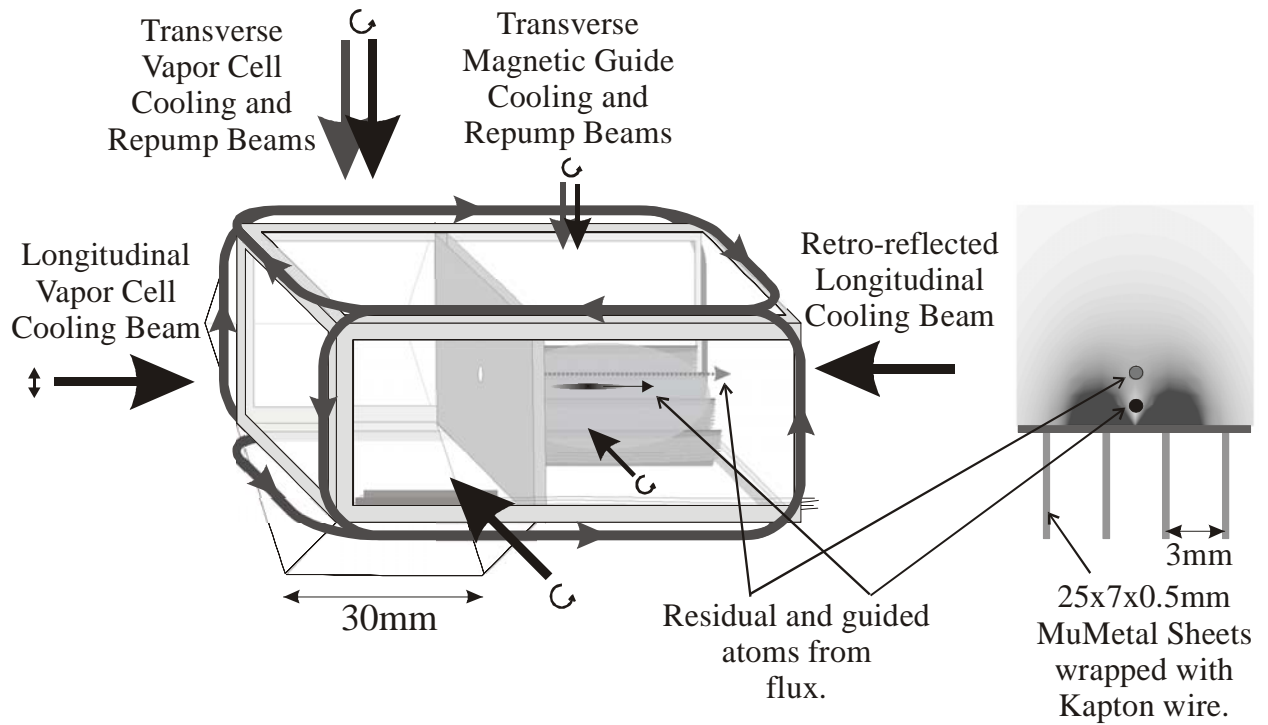

**Figure 1 – Schematic of the apparatus used, comprising of the vapor cell MOT to the left of the central gold mirror and the magnetic guide to the right. The external size of the cell was 70x30x30mm and mounted horizontally on a vacuum system flange. The right inset shows the axial view of the flux entering the field of the magnetic guide and the position of the guide minimum.**

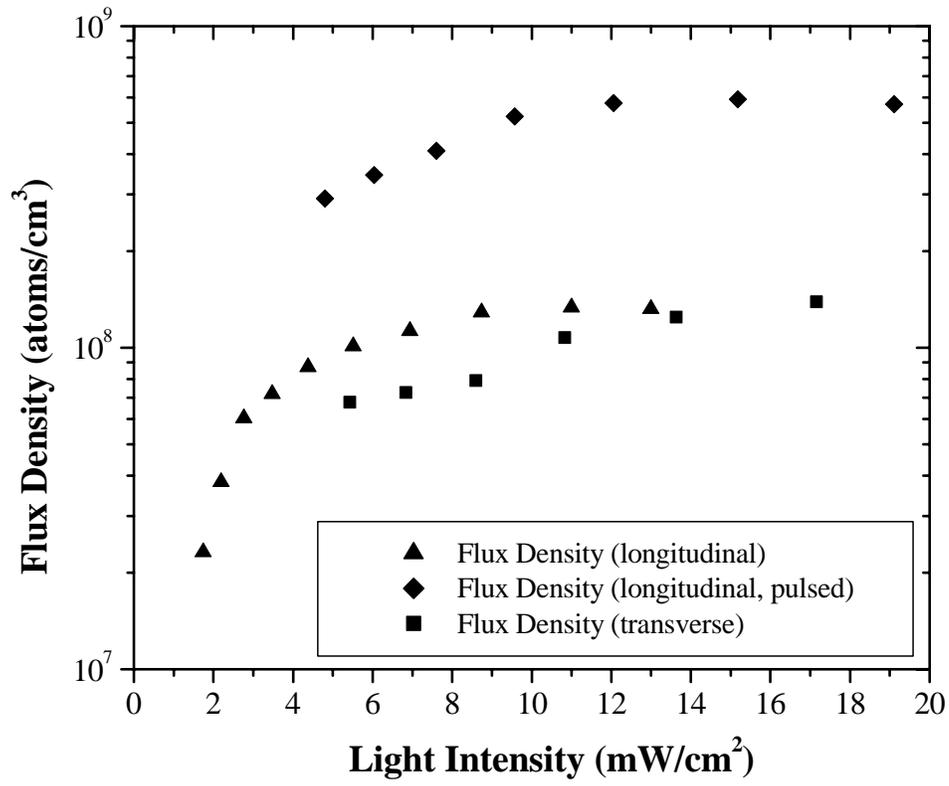

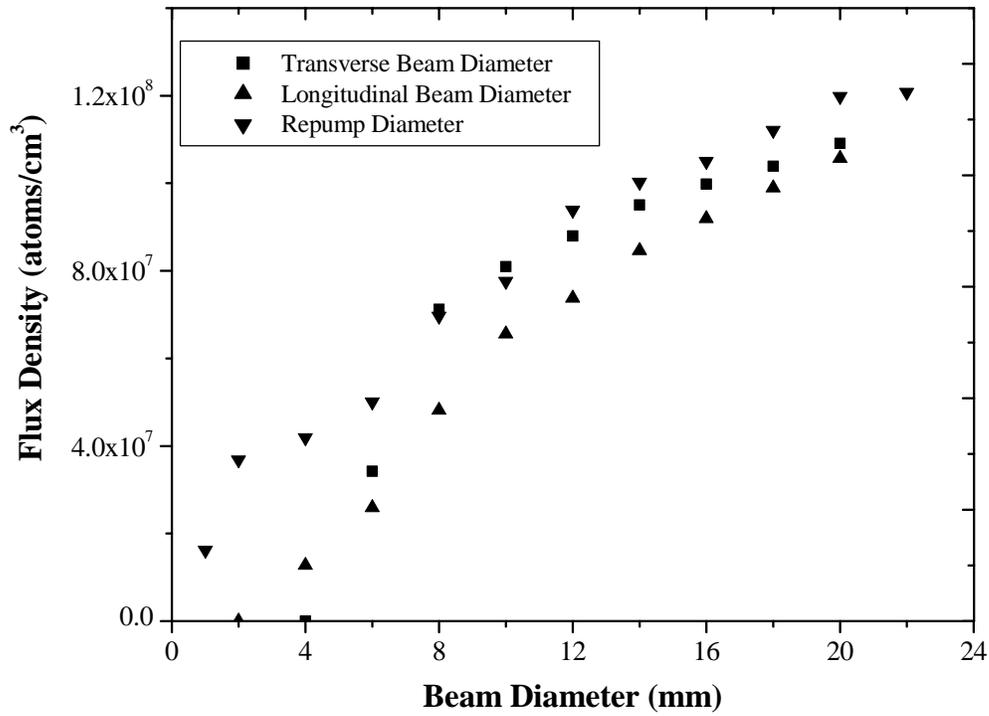

**Figure 2 – Flux density of the atoms exiting the hole as a function of a) the transverse and longitudinal light intensities and b) the diameter of the cooling and repumping beams.**

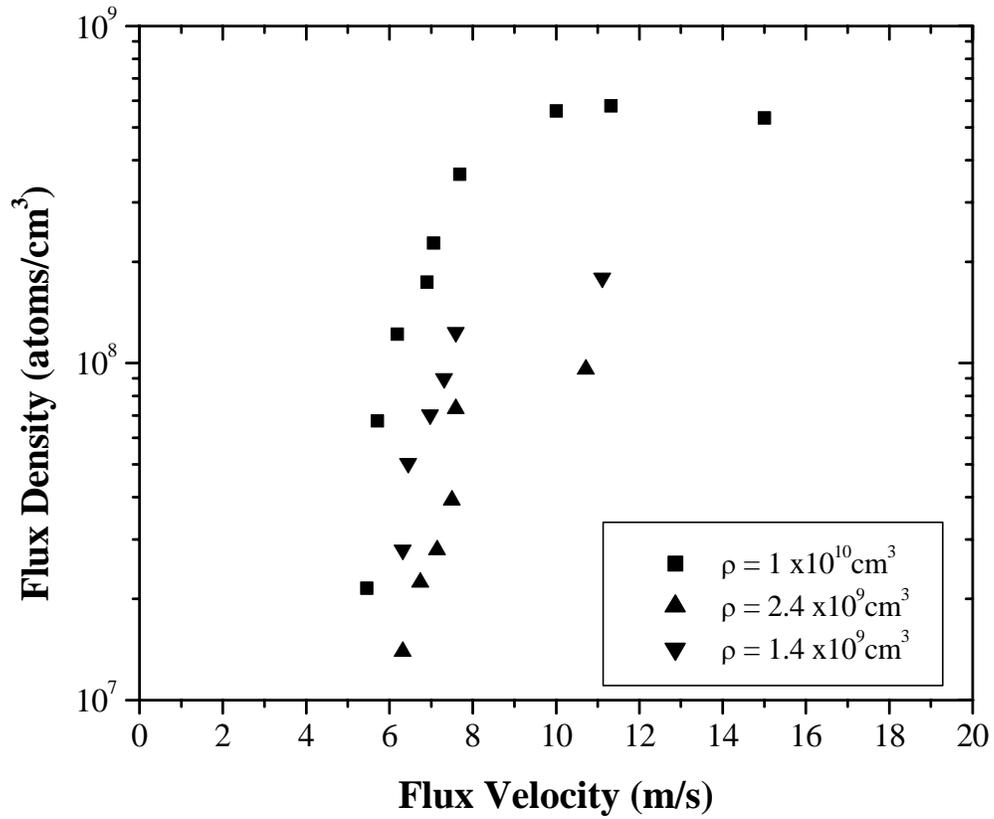

**Figure 3 – Flux density as a function of the flux velocity after the hole, controlled by the intensity of the retarding beam, for different background rubidium densities.**

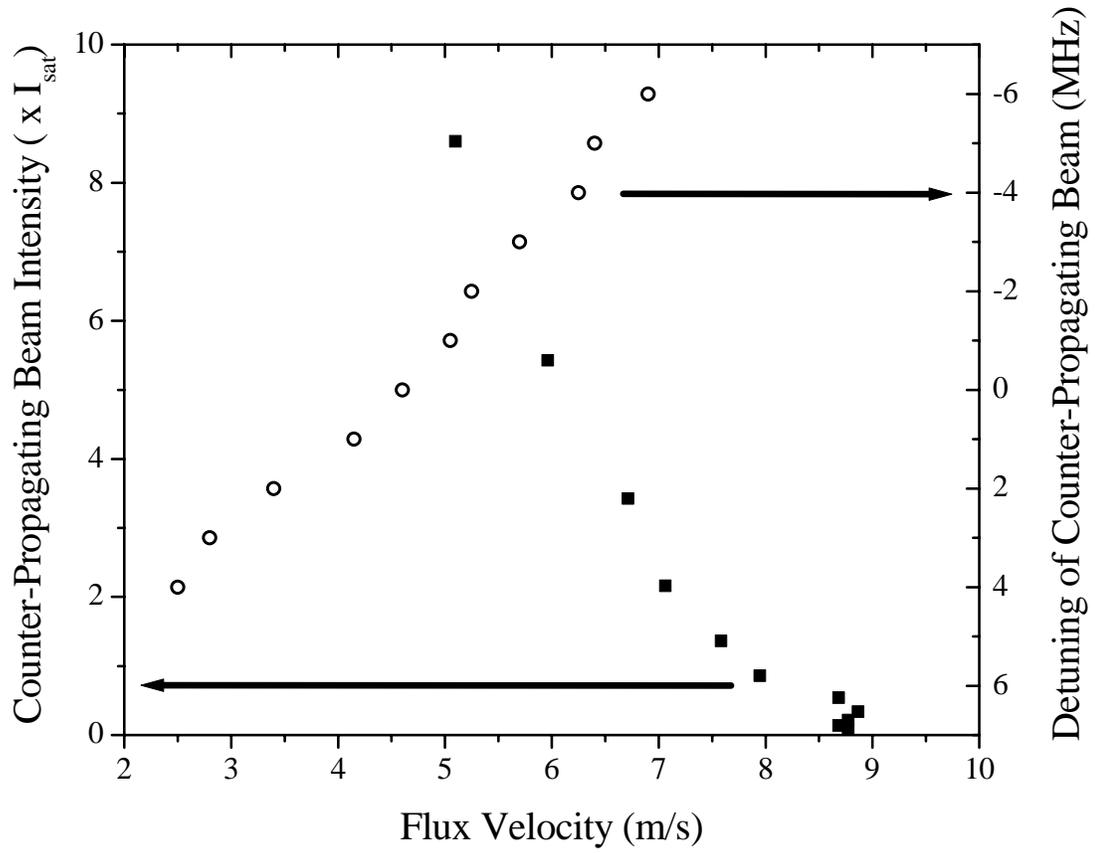

**Figure 4** –Velocity of the flux after the hole as a function of the intensity and detuning of the retarding beam with respect to the pushing beam.